\documentclass[%
 reprint, 
 amsmath,amssymb,
 aps, physrev,
]{revtex4-2}
\usepackage{hyperref}
\usepackage[left=2.0cm,right=2.0cm,top=2.5cm,bottom=2.5cm]{geometry}
\usepackage{graphicx}
\usepackage{amssymb,amsmath,amscd}
\usepackage{mathtools, color}
\usepackage{cleveref}
\usepackage{multirow}

\usepackage{xcolor}
\usepackage{dcolumn}
\usepackage{bm}

\usepackage[utf8]{inputenc}
\usepackage[T1]{fontenc}
\usepackage{mathptmx}
\usepackage{etoolbox}

\begin{document}

\title{Periodic and quasiperiodic traveling waves in nonlinear lattices with odd elasticity}







\author{Andrus Giraldo}%
 \email{agiraldo@kias.re.kr}
\affiliation{School of Computational Sciences, Korea Institute for Advanced Study, Seoul, South Korea
}
\author{Stefan Ruschel}%
 \email{stefan.ruschel@nottingham.ac.uk}
\affiliation{School of Mathematical Sciences, University of Nottingham, United Kingdom
}
\author{Behrooz Yousefzadeh}%
 \email{behrooz.yousefzadeh@concordia.ca}
\affiliation{Department of Mechanical, Industrial \& Aerospace Engineering, Concordia University, Montreal, QC, Canada
}



\begin{abstract}
Discrete nonlinear systems support a rich variety of localized and extended wave phenomena, with their dynamics sensitively dependent on the symmetries of the underlying interaction forces within the lattice. Odd elasticity, emerging in effective models of active materials, breaks the action-reaction symmetry of the local interactions and gives rise to new wave behavior. We investigate the existence and stability of traveling waves in a nonlinear lattice with odd elasticity, where the coupling force between adjacent units depends asymmetrically on the deformations of the coupled units (nonreciprocal elastic coupling). We demonstrate the existence of periodic and quasiperiodic traveling waves and analyze their spectral stability using the master stability framework. In particular, we identify the onset of Eckhaus instability based on the curvature of the associated master stability curve. This approach enables a quantitative analysis of size effects, specifically the bounds on lattice sizes for which a given traveling wave is stable. The stability analysis for quasiperiodic waves is based on an effective description of the envelope of the response through a rotating wave approximation, which agrees well with direct numerical simulations. Our findings establish a unified framework for understanding wave propagation characteristics in nonlinear lattices, for both periodic and quasiperiodic wave profiles. We highlight, qualitatively and quantitatively, the role of nonreciprocal stiffness on the existence and stability of nonlinear traveling waves in dissipative systems, and discuss how localization and stability depend on the interplay between nonlinearity, dissipation and odd elasticity. 

\end{abstract}

\maketitle


\section{Introduction}

Controlling the flow of mechanical wave energy is an essential feature in many technologies, with applications ranging from vibration mitigation and impact protection~\cite{transferPath,seismic} to energy-harvesting~\cite{harvestingAndrea}, resonant-based sensing~\cite{MEMSsensing} and mechanical computing~\cite{computingReview}. Reciprocity theorems have played a key role in our understanding, analysis, and design of wave control devices and materials since the 19th century~\cite{rayleigh,lamb,cook}. They ensure that the transmission characteristics between two points remain invariant when the source and receiver are interchanged. More recently, research on nonreciprocity has emerged from the technological need for devices and materials that can enable directional control over mechanical energy flow and the quest for expanding the range of available wave-based functionalities, including vibration isolation and impact mitigation~\cite{nonreciprocalIsolation1,nonreciprocalIsolation2}, energy harvesting~\cite{TET}, asymmetric energy amplification~\cite{nonreciprocalAntonio}, directional signal processing~\cite{computingMostafa} and mechanical locomotion~\cite{coulaisLocomotion}.

Nonreciprocity is most often (though not exclusively) associated with nonlinear or active systems~\cite{BYnonreciprocity,NRMsymmetry}. In nonlinear systems, breaking the mirror symmetry of the medium is a necessary (but not sufficient~\cite{GiraldoYousefzadeh}) condition for enabling nonreciprocal dynamics. In active systems, two common means of enabling nonreciprocity involve spatiotemporal modulation of the effective properties of the system (parametric excitation)~\cite{husseinPRSA} and odd elasticity~\cite{oddElasticity}. 

Materials with odd elasticity possess anti-symmetric components in their elastic tensor. One consequence of this asymmetry is that the work required for a quasistatic deformation of these materials depends on the deformation path, as opposed to only the initial and final states. Energy is no longer conserved in these materials, and their physical realization requires an external source of energy~\cite{coulaisOddLinear,guoliangNatureComm,guoliangPNAS,sirota2025JASA}. Thus, materials with odd elasticity are active materials. 

\begin{figure*}
    \centering
\includegraphics[width=1.\linewidth]{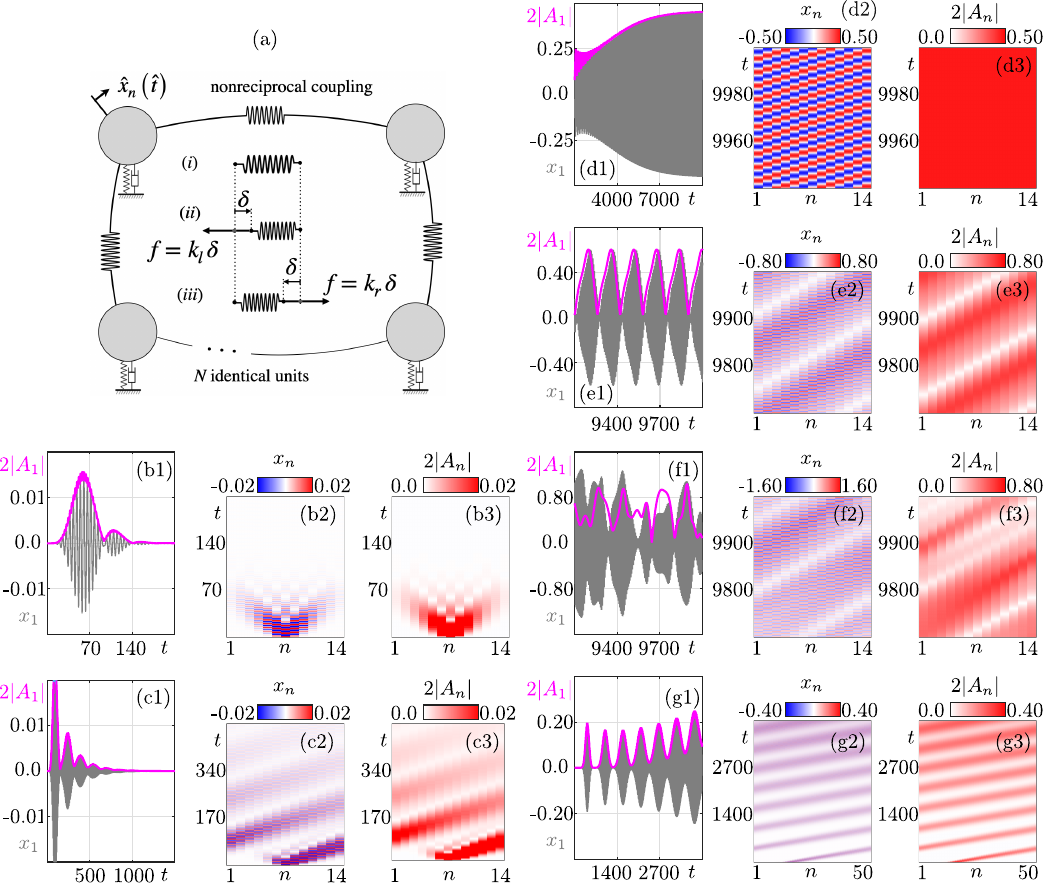}
    \caption{Wave propagation in a nonlinear lattice with odd elasticity. Panel~(a) shows a schematic of the mechanical system~\eqref{eq:dimension}  under consideration. Other panels present numerical simulations of (non-dimensionalized) systems~\eqref{eq:main} and~\eqref{eq:envMainFreq}, which govern oscillator displacements and the evolution of the corresponding envelope, respectively. More specifically, panels~(b1), (c1)--(g1) display the displacement of the first oscillator (gray curve) alongside the solution of the envelope equation (purple curve). Remaining panels show spatiotemporal plots of the displacement $x_n$, and the envelope $2A_n$, respectively . In panels~(b)–(f), the  size of the lattice is $N = 14$, while panel~(g) corresponds to $N = 50$. System parameters are in panel~(b), $(\alpha,\omega_0) = (0, 1)$; in panel~(c), $(\alpha,\omega_0) = \bigl(0.5, -\alpha\kappa/\zeta \sin(6\pi/N)\bigr)$; in panel~(d), $(\alpha,\omega_0) = \bigl(0.555, -\alpha\kappa/\zeta \sin(6\pi/N)\bigr)$; in panel~(e), $(\alpha,\omega_0) = \bigl(0.58, -\alpha\kappa/\zeta \sin(6\pi/N)\bigr)$; in panel~(f), $(\alpha,\omega_0) = \bigl(0.61, -\alpha\kappa/\zeta \sin(2\pi/N)\bigr)$; and in panel~(g), $(\alpha,\omega_0) = (0.55, 1)$. Other parameters are $(\zeta,\kappa,\beta) = (0.05, 0.1, 0.1)$. The initial conditions are the same in panels (b) and (c).}
    \label{fig:1}
\end{figure*}

A mechanical example of odd elasticity is a nonreciprocal spring. For a given deformation, a nonreciprocal spring produces a force of different magnitude depending on which of its ends is deformed; see Fig.~\ref{fig:1}(a). In mechanical systems with nonreciprocal springs, unidirectional transmission may occur at all propagation frequencies~\cite{coulaisOddLinear}. A balanced interplay between odd elasticity (spring nonreciprocity) and mechanical damping in these structures allows for sustained free wave propagation in a very limited parameter range~\cite{nonreciprocalRajesh}. Disturbing this balance results in exponential growth of amplitude in one direction and decay in the other -- this is what linear theory predicts, in contrast to the bounded amplitude observed in experimental observations~\cite{coulaisOddNonlinear}. The nonlinear restoring forces that appear in experiments at higher amplitudes provide the balance required for long-lived nonlinear waves to exist and travel through the lattice. The role of nonlinearity in sustaining non-decaying traveling waves in a system with local nonreciprocal interaction has already been reported in both discrete and continuum systems~\cite{coulaisOddNonlinear,coulaisSoliton,jana2025harnessing,sandoval2025non}. Localized modes have been reported as non-Hermitian skin modes in a long lattice~\cite{manda2025skin} and as vibration modes of coupled mechanical oscillators~\cite{nonreciprocalDuffingNNM}. The active nature of nonreciprocal interaction allows damped oscillators to sustain chaotic motion in the absence of direct external drive~\cite{nonreciprocalDuffingIJNM}.

It is important to disambiguate two different notions of nonreciprocity before proceeding. One notion refers to asymmetric propagation of waves (information) along the same transmission channel. In this sense, it is the response of the system, not the system itself, that is nonreciprocal. The other notion refers to the symmetry of the interaction between two points, for example through a spring connecting two masses. A nonreciprocal spring is one through which forces of different magnitudes are transferred depending on the direction of the input. This is a counterintuitive behavior for a mechanical spring because it violates Newton's third law of motion. It is relevant here to clarify that a bilinear spring, one with different constants in compression and extension, is reciprocal. 

In this work, we explore the combined effects of nonlinearity and odd elasticity on the existence and stability of traveling waves through a discrete model of a material with local nonreciprocal interaction. Because of nonreciprocal coupling forces, all the wave phenomena that we investigate are also nonreciprocal in terms of direction of propagation. As a prototypical system, we consider a ring of $N$ identical oscillators with cubic stiffness nonlinearity that are coupled via springs with nonreciprocal stiffness. This system is modeled as a chain of Duffing oscillators with periodic boundary conditions; see system~\eqref{eq:main} and Sec.~\ref{sec:system} for details, and Fig.~\ref{fig:1}(a) for a schematic of the system. The nonreciprocity implies that the restoring force depends on which end of the spring is deformed.  

The primary objective of the present work is to provide a systematic analysis of periodic and quasiperiodic traveling waves in a nonlinear lattice with nonreciprocal coupling stiffness.
We use analytical and numerical methods to provide the conditions for existence and stability of periodic traveling waves as a function of the degree of stiffness nonreciprocity. Furthermore, we highlight the emergence of quasiperiodic traveling waves by means of secondary instabilities (torus bifurcations) and employ a rotating-wave approximation to analyze the (anharmonic) envelope of the quasiperiodic traveling waves and their instabilities. 

We adopt the recently developed master stability framework for traveling waves on networks and lattices~\cite{RuschelGiraldo}. This allows for a systematic analysis of the existence and stability of quasiperiodic traveling waves independently of the network size by introducing a co-moving coordinate frame for perturbations along the wave. The ensuing master stability curves are computed using numerical continuation techniques. Most notably, we compute ---for both periodic and quasiperiodic traveling waves--- the size-independent stability boundary defined by an Eckhaus curve \cite{eckhaus2012studies, TuckermanBarkley,hoyle2006pattern, cross1993pattern}.
In the case of quasiperiodic waves, we achieve this by analyzing the envelope of the response, i.e., periodic traveling waves in the rotating-wave approximation.
The numerical methods are validated by direct comparison of the spectrum and stability boundary of periodic traveling waves with rotating waves in the approximation. 

Section~\ref{sec:system} introduces the mathematical model of the nonlinear system with nonreciprocal coupling that we investigate in this work. We review propagation of linear waves in a system with nonreciprocal coupling in Section~\ref{sec:nonlinearPlaneWave}. We investigate the birth of traveling waves through a Hopf bifurcation and provide a detailed stability analysis based on the master stability framework. Specifically, we discuss the influence of the lattice size on wave stability and determine the stability boundaries based on the onset of Eckhaus instability. In Section~\ref{sec:anharmonic}, we extend the analysis to quasiperiodic traveling waves. We use the rotating-wave approximation to derive the envelope equations. We then investigate the conditions for the existence and stability of quasiperiodic traveling waves using the master stability framework. We provide details of the master stability framework in Section~\ref{sec:compEck}. We conclude the paper in Section~\ref{sec:conclusions} by reviewing our main findings and suggestions for future investigations motivated by this work. 

\section{Mathematical model}
\label{sec:system}

We consider a discrete model of a mechanical system with nonreciprocal coupling springs, with a schematic representation in Fig.~\ref{fig:1}(a). In terms of physical parameters, the governing equations are
\begin{equation}
    \label{eq:dimension}
    m\ddot{\hat x}_n + c\dot{\hat x}_n+ k_g \hat x_n +k_r(\hat x_n-\hat x_{n+1})+k_l(\hat x_n-\hat x_{n-1})  + k_{f} \hat x_n^3= 0
\end{equation}
Here, $\hat x_n (\hat t\,)$,  $n=1,...N$, denotes the displacement of $n$-th degree of freedom (with mass $m$) from its static equilibrium position, overdot denotes differentiation with respect to time, $c$ is the coefficient of viscous damping, $k_g$ represents a spring that connects each mass to the ground, $k_r$ is the spring constant on the right-hand side of each mass, $k_l$ is the spring constant on the left-hand side, and $k_{f}$ is the coefficient of the nonlinear spring. 

In a physical setup, the nonlinear stiffness terms are determined by fitting a truncated Taylor expansion to measurements; {i.e.} a system identification procedure. In that spirit, our adoption of a cubic stiffness coefficient implies that the quadratic stiffness is relatively negligible. This is typical of many mechanical systems with a symmetric restoring force, such as geometric nonlinearity in cantilever beams~\cite{duffing}. In the absence of a physical setup, ignoring the quadratic term allows us to focus on the most typical effects of nonlinearity.

To non-dimensionalize the equations, we rescale the displacements with a suitable length scale $d$ such that $x=\hat x/d$ is non-dimensional. In experiments, $d$ often represents a physical length such as the distance between adjacent masses at static equilibrium (the lattice constant). We rescale time such that $t=\omega_0\hat t$ is the non-dimensional time, with $\omega_0=\sqrt{k_g/m}$. For the nonreciprocal coupling springs, we assume that $k_l=k_c(1+\alpha)$ and $k_r=k_c(1-\alpha)$. Dividing through by $k_g$, we arrive at the following non-dimensional equations
\begin{eqnarray}
    \label{eq:main}
    \ddot{x}_n + 2\zeta\dot{x}_n+ x_n + \kappa (2x_n-x_{n+1}-x_{n-1}) \\ \nonumber
    + \alpha \kappa (x_{n+1}-x_{n-1}) 
    + \beta x_n^3= 0
\end{eqnarray}
Here, $\zeta=c/2\sqrt{mk_g}$ is the damping ratio, $\kappa=k_c/k_g$ is the non-dimensional coupling coefficient and $\beta=k_{f}d^2/k_g$ is the non-dimensional coefficient of nonlinearity. 

Parameter $\alpha$ represents the degree of nonreciprocity in the spring constant (odd elasticity), with $\alpha=0$ corresponding to a reciprocal connection (regular mechanical spring) and $0<\alpha<1$ corresponding to a nonreciprocal connection. When $\alpha=1$, each mass only feels the spring on its right-hand side, and the spring constant on the left becomes negative when $\alpha>1$. We only consider $0\le\alpha\le1$ in this work.

Throughout the manuscript, we restrict our analysis to the case $\zeta\geq0$. Additionally, given the ring arrangement of the lattice of $N$ oscillators, we can restrict to $\alpha\geq 0$ because system~\eqref{eq:main} is invariant under the simultaneous actions of transformations $n\mapsto N-n+1$ and $\alpha\mapsto -\alpha.$ 

Before proceeding to technical details, we illustrate typical behavior of the system by numerically computing the time series of the non-dimensionalized model \eqref{eq:dimension}. Figs.~\ref{fig:1}(b)-(g) showcase representative dynamical regimes of these wave phenomena. In addition, we present the evolution of the envelope of these solution trajectories by simulating the corresponding envelope equation obtained by the rotating wave approximation ~\eqref{eq:envMainFreq}; see Sec.~\ref{sec:anharmonic} for further details.

Figs.~\ref{fig:1}(b)-(c) highlight the role of nonreciprocal coupling in enabling nonreciprocal propagation. Panel (b2) shows the spatiotemporal evolution of an initial displacement applied to one unit in a lattice of 14 oscillators for a system with reciprocal coupling ($\alpha=0$). The disturbance travels in two opposite directions in an identical manner, as expected. Panel (c2) shows the evolution of the same disturbance in a system with nonreciprocal coupling ($\alpha=0.5$). The disturbance persists for a longer duration and, remarkably, propagates in only one direction. This unidirectional nature of wave propagation in systems with nonreciprocal coupling persists in the presence of nonlinearity, as shown in the spatiotemporal plots of panels (d)-(g) and as observed in experiments~\cite{coulaisOddLinear}. 

Fig.~\ref{fig:1}(d) shows an example of a periodic traveling wave. Panel~(d1) shows the time-domain response for one oscillator as the transient response decays -- both the displacement and its envelope are shown. Panel~(d2) shows the spatiotemporal evolution of the emerging traveling wave in the steady state. Panel~(d3) shows the spatiotemporal evolution of the response envelope over the same time period as in panel~(d2). Because the traveling wave in panel~(d2) is periodic, its envelope has a constant value (indicating the amplitude of the wave). Thus, panel~(d3) shows no variation. 

Fig.~\ref{fig:1}(e) shows a quasiperiodic traveling wave. Note in panel~(e1) that the response envelope is periodic and anharmonic. Accordingly, the spatiotemporal plot of the envelope shows a periodic traveling wave with the speed of the slow oscillations of the quasiperiodic motion. Such quasiperiodic waves have already been observed in experiments~\cite{coulaisOddNonlinear}. We will use the periodic nature of the response envelope to analyze the existence and stability of quasiperiodic traveling waves. 

Fig.~\ref{fig:1}(f) shows a traveling wave that appears to be chaotic. This indicates that the destabilization of the response envelope coincides with the destabilization of the underlying response. We also observe that our envelope equations do not capture the response envelope accurately in case of chaotic motion. Note that propagation remains nonreciprocal.

Fig.~\ref{fig:1}(g) shows an example of a quasiperiodic traveling wave with a spatially localized waveform; {cf.} panel~(e). Such localized traveling waves start to appear as the number of units on the lattice increases because more localized solutions have a lower wavenumber. However, these localized waveforms are unstable and can only be observed over a finite time interval.

\section{Periodic traveling waves}
\label{sec:nonlinearPlaneWave}

\subsection{Linear analysis and dispersion relation}
\label{sec:linearWaves}

In a linear system, the dispersion relation describes how waves of different frequencies propagate through the medium. To obtain the (complex-valued) dispersion relation of system~\eqref{eq:main} in the absence of nonlinearity, we set $\beta=0$ and substitute the plane-wave ansatz $x_n(t)=A\exp\left(i(\omega t-qn)\right)$ with $\rm{Re}(\omega)\geq 0$ and $q\in(-\pi,\pi]$ into system~(\ref{eq:main}) to obtain the consistency relation
\begin{equation}\label{eq:plane-wave-consistency}
    \omega^2 -2i\omega\zeta-\left(1 + \kappa (2-2\cos(q)) +2i\alpha\kappa\sin(q) \right)=0.
\end{equation} 
The parameters $A$, $\omega$, and $q$ correspond to the amplitude, (angular) frequency, and (angular) wave number of the plane wave, respectively.
Solving \eqref{eq:plane-wave-consistency} for $\omega$ as a function of $q$, we obtain the (complex-valued) dispersion relation of system~\eqref{eq:main}, which can be compactly written as
\begin{equation}\label{eq:dispersion}
    \omega(q) = i\zeta +\sqrt{r(q)}e^{i\theta(q)/2}, 
\end{equation}
where $r(q)\geq 0$ and $\theta(q)\in[0,2\pi)$ satisfy $r(q)\exp(i\theta(q))=1+\kappa(2-2\cos(q))+\zeta^2-2i\alpha\kappa\sin(q)$. We have chosen the positive root of $r(q)\exp(i\theta(q))$ to ensure $\rm{Re}(\omega)\geq0$.
Thus, the real part of the frequency, $\rm{Re}(\omega)=\sqrt{r}\cos(\theta/2)$, contributes to oscillatory motion in time and its imaginary part, $\rm{Im}(\omega)=\zeta\pm\sqrt{r}\sin(\theta/2)$, contributes to exponential decay ($\rm{Im}(\omega)>0$) or growth ($\rm{Im}(\omega)<0$) of a given plane wave with wavenumber $q$.

In the absence of dissipation and nonreciprocity ($\zeta=0$, $\alpha=0$), Eq.~(\ref{eq:plane-wave-consistency}) simplifies to the well-known dispersion relation $\omega^2=1+4\kappa\sin^2(q/2)$ for a monatomic lattice~\cite{AMR2014}. 
For $\zeta>0$ and $\alpha=0$ (reciprocal springs), we have $\theta=0$ and $\omega=i\zeta+\sqrt{r}$; therefore, plane waves decay exponentially in both directions. 

\begin{figure}
    \centering\includegraphics[width=\linewidth]{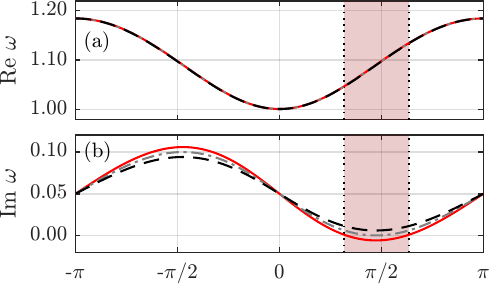}
    \caption{Dispersion relation of system~\eqref{eq:main} (neglecting nonlinearity) for different values of $\alpha$. Shown are the real part (a) and imaginary part (b) of $\omega$ as defined in \eqref{eq:dispersion} as a function of wave number $q$. Each panel shows three curves for different values $\alpha=0.49$ (black, dashed),  $\alpha=0.545$ (gray, dashed-dotted), and $\alpha=0.61$ (red, solid) -- the three curves coincide visually in panel (a). The region of wavenumbers $q$ with $\rm{Im}(\omega)<0$  (highlighted with red background) corresponds to wave numbers that result in exponential amplitude growth for $\alpha=0.61$ (red, solid). The critical values of $q$ with $\rm{Im}(\omega)=0$ are highlighted by black dotted lines for $\alpha=0.61$. Other system parameters are $\zeta=0.05$, $\kappa=0.1$, $\beta=0$.}
    \label{fig:dispersion}
\end{figure}

For $\alpha\ne0$ (nonreciprocal spring), it is possible to find a balance between damping and nonreciprocity that eliminates the imaginary part of the propagation frequency. Thus, it becomes possible for waves to propagate through the lattice without decay~\cite{coulaisOddLinear,nonreciprocalRajesh}. 
Figure \ref{fig:dispersion} shows the real part (a) and the imaginary part (b) of the dispersion curve \eqref{eq:dispersion} for $\alpha=0.49$ (black, dashed),  $\alpha=0.545$ (gray, dashed-dotted), $\alpha=0.61$ (red, solid), and fixed parameter values $\zeta=0.05$, $\kappa=0.1$, $\beta=0$.  
Panel (b) shows that $\alpha$ renders the imaginary part of the frequency dependent on the wavenumber. 
In particular, increasing $\alpha$ leads first to a critical condition for non-decaying plane waves ($\rm{Im}(\omega)=0$), and subsequently to the appearance of a range of wavenumbers over which the corresponding plane waves grow in amplitude ($\rm{Im}(\omega)<0$); {cf.} red shaded region in Fig.~\ref{fig:dispersion}.

The condition $\rm{Im}(\omega(q))=0$ is equivalent to a critical value of $\alpha=\alpha_H(q),$ where
\begin{equation}
    \label{eq:dispersion-zero-growth}\alpha_H(q)=\frac{\zeta\sqrt{1+4\kappa\sin^2(q/2)}}{\kappa \sin(q)}, 
\end{equation} 
This is obtained by separating \eqref{eq:plane-wave-consistency} into its real and imaginary parts and solving for $\alpha\geq0$ for a real-valued, non-negative propagation frequency. 

\begin{figure}
    \centering\includegraphics[width=\linewidth]{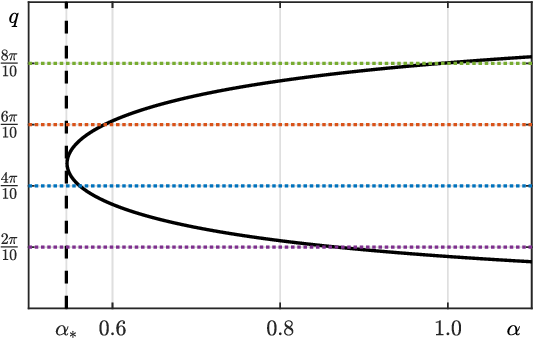}
    \caption{Critical curve $a_H(q)$ (solid) in the $(\alpha,\kappa)$-plane for parameter values $\zeta=0.05$, $\kappa=0.1$, $\beta=0$. The vertical dashed line shows the locus of $\alpha_\ast=\min_q a_H(q)\approx 0.545$. Horizontal dotted lines indicate intersection of the critical curve with wave numbers $q=2\pi/10,\ldots,8\pi/10.$}
    \label{fig:critical-curve}
\end{figure}

Figure \ref{fig:critical-curve} shows $\alpha_H$ as a function of $q$. The curve $a_H(q)$ is strictly convex for $-\pi\leq q\leq \pi$ and attains its minimum $\alpha_\ast=\min_q a_H(q)$ at $q=q_\ast,$ where
$$\cos(q_\ast)=\frac{1+2\kappa-\sqrt{1+4\kappa}}{2\kappa},$$
with value 
$$\alpha_\ast = \frac{|\zeta|\sqrt{1+2\kappa+\sqrt{1+4\kappa}}}{\sqrt{2\kappa^2}}$$.

For all $\alpha<\alpha_\ast$, any plane wave will decay exponentially irrespective of its wave number; cf. Fig. \ref{fig:dispersion}(b). At the critical value $\alpha=\alpha_\ast$, the energy loss through damping balances the energy gain through the nonreciprocal spring (an active element) to allow free propagation of plane waves through the system with wavenumber $q=q_\ast$. For $\alpha>\alpha_\ast,$ there exists wave numbers $q_1,q_2\in[-\pi,\pi)$ such that the corresponding plane waves neither grow nor decay, while all plane waves with wave numbers $q_1< q < q_2$ grow exponentially; {cf.} Fig.~\ref{fig:dispersion}(b) where $q_1\approx1$ and $q_2\approx2.$  
Note again that these plane waves are induced by an interplay of damping and nonreciprocity. In the absence of damping ($\zeta=0$), the imaginary part of $\omega$ vanishes only at $\sin q=0$, which corresponds to waves with zero group velocities (standing waves).

\subsection{Hopf bifurcation: Onset of nonlinear traveling waves}

While one can use the results of the linear analysis to determine the onset of oscillatory instabilities in the nonlinear system, we recast these results in a dynamical systems framework, which provides the notation and tools used in what follows. From this perspective, the onset of nonlinear (finite-amplitude) traveling waves corresponds to a Hopf bifurcation, i.e., the onset of nonlinear (finite-amplitude) traveling waves corresponds to a Hopf bifurcation. This onset corresponds to a balance between nonreciprocity $\alpha$ and damping $\zeta$. For a damped reciprocal system, any perturbation of the static equilibrium decays, which means that $x_n=0$ is a stable equilibrium. As the value of $\alpha$ gradually increases, it reaches a critical value at which perturbations of the equilibrium result in periodic motion; {i.e.} traveling waves through the lattice.

To predict the onset of traveling waves, we study the stability of perturbations $v_n(t)$ around the equilibrium point $x_n=0$. Linearizing system~(\ref{eq:main}) at this equilibrium gives
$$\ddot{v}_n + 2\zeta\dot{v}_n+ v_n + \kappa (2v_n-v_{n+1}-v_{n-1}) + \alpha \kappa (v_{n+1}-v_{n-1})  = 0$$
Introducing the variables $w_n=\dot v_n$ and $z_n=(v_n,w_n)$, we can rewrite the equation for the evolution of perturbations as a first-order system
\begin{eqnarray*}
    \label{eq:linZero}
    &\dot z_n = \\ \nonumber
    &\begin{pmatrix}
        w_n \\
        -2\zeta w_n - v_n - \kappa (2v_n-v_{n+1}-v_{n-1}) - \alpha \kappa (v_{n+1}-v_{n-1})
    \end{pmatrix}
\end{eqnarray*}
We apply a Floquet-Bloch decomposition $\hat{z}_q=\sum_{n\in \mathbb{Z}} e^{-iqn}z_n$, where each subindex $q$ represents a mode (a spatial Fourier mode). Thus, each wave component evolves according to the following diagonalized system of equations
\begin{equation*}
\label{eq:linZeroFloqBloch}
    \dot{\hat{z}}_q =
    \begin{pmatrix}
    0 & 1 \\
    -1-2\kappa(1-\cos(q))-2i\alpha \kappa \sin(q) & -2\zeta 
    \end{pmatrix}\begin{pmatrix}
        \hat{v}_q \\
        \hat{w}_q
    \end{pmatrix}.
\end{equation*}
The real parts of the eigenvalues of this matrix determine the stability of perturbations. The characteristic equation of the zero equilibrium is given by
$$\lambda^2+2\zeta \lambda +1+2\kappa(1-\cos(q))+2i\alpha \kappa \sin(q)=0.$$
To capture the onset of the Hopf bifurcation, we set $\lambda =i\nu$. This results in a system of algebraic equations for $\nu$ and $q$,
\begin{eqnarray}
\label{eq:NAG1}
    &\nu^2&=1+2\kappa (1-\cos(q)) \\
\label{eq:NAGtemp}
    &0&=\nu \zeta+\alpha \kappa \sin(q)
\end{eqnarray}
Eq.~(\ref{eq:NAG1}) relates the frequency of the traveling waves (periodic solutions borne from the Hopf bifurcation) to the wavenumber $q$ and the coupling coefficient $\kappa$. The system \eqref{eq:NAG1}--\eqref{eq:NAGtemp} recovers the real and imaginary parts of \eqref{eq:plane-wave-consistency}. 
Eqs.~(\ref{eq:NAG1}) and (\ref{eq:NAGtemp}) can be combined to yield an implicit relation for the locus of the Hopf bifurcation, 
\begin{equation}
\label{eq:NAG2}
    \zeta^2(-1-2\kappa+2\kappa\cos(q))+ \alpha^2\kappa^2\sin^2(q)=0
\end{equation}
For a given $\zeta$ and $\kappa$, Eq.~(\ref{eq:NAG2}) describes the locus of the Hopf curve in the $(\alpha,q)$-plane. 
By taking the partial derivative with respect to $q$, we find that the minimum value for the onset of Hopf bifurcation (traveling waves) is given by $q_*$ and $\alpha_*$.

\begin{figure}
    \centering
    \includegraphics[width=\linewidth]{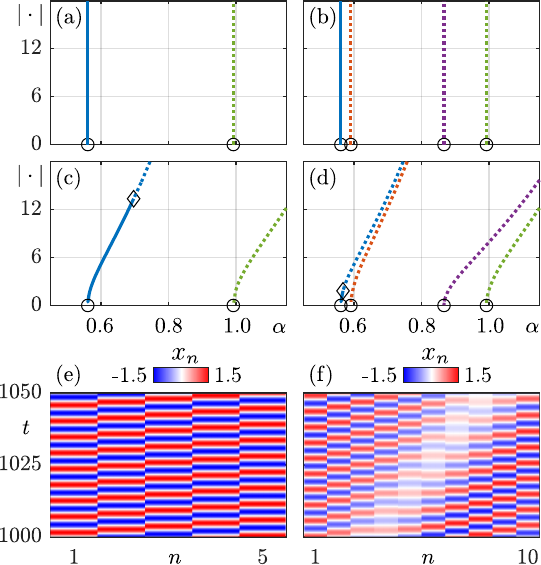}
    \caption{Numerical bifurcation diagram of periodic traveling waves in $\alpha$, showing the $L_2$-norm of the periodic profile, for $\beta=0$ (a)--(b) and $\beta=0.1$ (c)--(d) with $N=5$ (left column) and $N=10$ (right column). Points along solid lines correspond to (linearly) orbitally stable waves; points along dotted lines correspond to unstable waves. Colors indicate wave number $q=2\pi/10$ (purple), $q=4\pi/10$ (blue), $q=6\pi/10$ (orange), $q=8\pi/10$ (green); cf. Fig.~\ref{fig:critical-curve}. Circle markers indicate Hopf bifurcations and diamond markers indicate torus bifurcations. Panels (e)--(f) show long-term behavior of simulations starting from near the periodic traveling waves with wave number $q=2\pi/5$ at $\alpha=0.6$ with (e) $N=5$ and (f) $N=10$. Other parameters are $\zeta=0.05$, $\kappa=0.1$.}
    \label{fig:4}
\end{figure}

For a fixed lattice size $N$, the admissible wavenumbers are given by $q = 2\pi M/N$, with $M = 0,1,\dots,N-1$. A Hopf bifurcation for the lattice of size $N$ occurs when the horizontal line corresponding to one of these wavenumbers intersects the Hopf curve in Fig.~\ref{fig:critical-curve}. Figure \ref{fig:4} shows the influence of nonlinearity $\beta$ and the number of oscillators $N$ on the periodic traveling waves emerging from Hopf bifurcations (black circles). We consider $N=5$ and $N=10=5\cdot 2$ to compare two embeddings of a traveling wave in lattices of different sizes. The vertical lines in panels (a) and (b) indicate families of coexisting periodic traveling waves (periodic orbits) for the linear system ($\beta=0$), for $N=5$ and $N=10$, respectively. Two families of traveling waves exist for $N=5$ with wave numbers $q=2\pi/5=4\pi/10$ (blue) and $q=2\pi/5=8\pi/10$ (green), respectively, whereas four families of traveling waves exist for $N=10$ with wave numbers $q=2\pi/10$ (purple), $q=4\pi/10$ (blue), $q=6\pi/10$ (orange), $q=8\pi/10$ (green). In Fig.~\ref{fig:4}, solid curves indicate traveling waves that are (linearly) orbitally stable, and dashed curves indicate traveling waves that are orbitally unstable. 

Panels (c) and (d) show the existence of the nonlinear traveling waves emerging from the static equilibrium at critical parameter values at the Hopf bifurcation points (black circles) for weak nonlinearity ($\beta=0.1$). Note that the critical value of $\alpha$ at a Hopf bifurcation does not depend on the value of $\beta$; thus, the Hopf bifurcation points for the linear and nonlinear systems must coincide in panels (a)--(d). The correspondence between the families of linear and nonlinear traveling waves is clearly distinguishable in Fig.~\ref{fig:4}; namely, nonlinearity causes the nonlinear families of periodic orbits to "bend" in the bifurcation diagram, such that they exist beyond the bifurcation point (at least locally). Physically, this corresponds to the amplitude-dependent hardening effect of the grounding spring ($\beta>0)$.

More importantly, in contrast to the linear case ---where periodic solutions exist only at discrete parameter values--- the nonlinear system admits periodic traveling waves over a range of $\alpha$-values. The result is the emergence of branches of periodic traveling waves, arising from Hopf bifurcations, the first of which is supercritical. Note that the traveling waves can exhibit secondary instabilities in the nonlinear system depending on the size of the system, $N$. Most notably, in panel (d), the family of traveling waves with wavenumber $q=4\pi/10$ (blue) loses stability through a torus bifurcation (diamond marker). This has major consequences for the dynamics of the system because no stable periodic traveling waves exist for values of $\alpha$ larger than this critical point; instead, quasi-periodic traveling waves emerge as observed in simulation. 
We observe that the critical value of $\alpha$ at which the traveling wave loses stability depends on the lattice size, shifting from $\alpha \approx 0.7$ for $N=5$ to a value near the Hopf bifurcation point for $N=10$.

Panels (e) and (f) show the spatiotemporal evolution of nonlinear traveling waves at $\alpha=0.6$ starting from the periodic profile with wavenumber $q=2\pi/5$ for $N=5$ and $N=10$, respectively. As expected from the bifurcation diagrams in panels (c) and (d), the initial periodic profile remains stable for $N=5$, while it loses stability for $N=10$. 

\subsection{Master stability of traveling waves}
\label{sec:computation}

The stable traveling waves that emerge from the Hopf bifurcation remain stable as $\alpha$ increases, until they potentially lose stability through a secondary bifurcation, such as a torus bifurcation. The parameter value at which a given traveling wave destabilizes generally depends on its associated wave number $q$, as showcased in the previous subsection. The stability of each traveling wave is determined from its Floquet spectrum \cite{kuznetsov}, obtained by linearizing the dynamics about the time-periodic orbit over one temporal period. The corresponding monodromy matrix maps infinitesimal perturbations from one period to the next, and its eigenvalues, the Floquet multipliers, determine whether perturbations decay or grow. Orbital stability requires that all nontrivial multipliers lie strictly inside the unit circle, or equivalently, that all nontrivial Floquet exponents have negative real part.

To elucidate how the wave number $q$ and the nonreciprocity parameter $\alpha$ influence the existence and stability of traveling waves, we employ the master stability for traveling waves \cite{RuschelGiraldo}. 
In this framework, the stability of a traveling wave in a given lattice is related to the location of its associated master stability curves in the complex plane. These curves contain the complete spectrum of the Floquet exponents of the traveling wave, as well as the spectra of all embeddings of its wave profile in larger finite lattices. In fact, the master stability curve can be interpreted as the essential spectrum of the traveling wave in the infinite lattice under $\ell^2$ perturbations. The master stability curves are computed using numerical continuation of a suitable two-point boundary value problem (2PBVP) of a delay differential equation of the mixed type. See Section~\ref{sec:compEck} for a discussion in the specific context of the current work. 

\begin{figure}
    \centering
    \includegraphics[width=\linewidth]{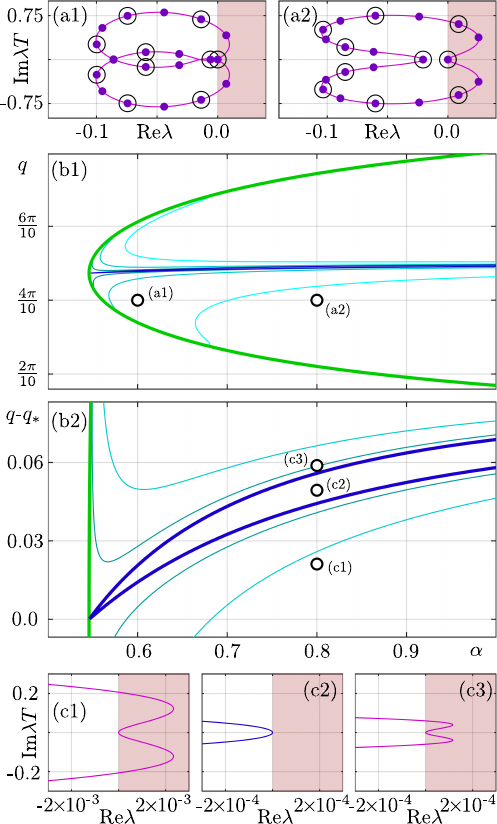}
    \caption{Panels~(a) shows the master stability curves of traveling waves in system~(\ref{eq:main}) for $(\alpha,q) = (0.6, 2\pi/5)$, panel~(a1), and $(\alpha,q) = (0.8, 2\pi/5)$, panel~(a2). Here, the dark circles and solid dots denote the Floquet exponents $\lambda$ for embeddings in lattices of sizes $5$ and $10$, respectively. Panel~(b1) shows the bifurcation diagram in the $(\alpha,q)$ parameter plane, including the Hopf bifurcation curve (green), loci of torus bifurcations for $N_{\min} = 5$ (cyan), $10$ (darker cyan), and $20$ (darkest cyan), and the Eckhaus instability curves (blue). Panel~(b2) presents an enlargement near the Eckhaus instability curves. Panels~(c1)–(c3) display the master stability curves near $\lambda = 0$ for traveling waves at the parameter values indicated in panel~(b2), namely $(q,\alpha) = (0.24, 0.8)$ in panel~(c1), $(0.2445, 0.8)$ in panel~(c2), and $(0.246, 0.8)$ in panel~(c3). In all  panels, $(\zeta,\kappa,\beta) = (0.05, 0.1, 0.1)$.}
    \label{fig:eckhaus-curve} 
\end{figure}

The wavenumber $q$ determines the admissible lattice sizes in which a given traveling wave can be embedded. In particular, if $q = 2\pi M_0/N_0$ with $M_0,N_0 \in \mathbb{N}$, then the traveling wave can be embedded in any lattice of size $N = m N_0$, with $m \in \mathbb{N}$, by repeating the waveform $m$ times; see \cite{RuschelGiraldo} for details. For convenience, we introduce the quantity $k=2\pi/q$, to facilitate the discussion in this subsection.

Panels~(a1) and~(a2) of Fig.~\ref{fig:eckhaus-curve} show the master stability curves associated with the traveling wave with $k = 1/5$ (wave number $q=2\pi/5$) at $\alpha = 0.6$ and $\alpha=0.8$, respectively. Because $k = 1/5$, the traveling wave can be embedded in lattices of size $5m$, with $m = 1,2,\ldots$. In particular, the black circle denotes the Floquet exponent $\lambda$ of the traveling wave for a lattice of size $5$ (its minimum admissible embedding), while the solid dots represent the corresponding Floquet exponents when the traveling wave is embedded in a lattice of size $10$, obtained by concatenating two copies of the wave. These Floquet exponents ---and those of any higher-dimensional embedding of the same wave profile--- lie on the same master stability curve. Moreover, at least one master stability curve necessarily passes through the origin of the complex plane, reflecting the zero Floquet exponent associated with time-periodic solutions. 

Notice in panel~(a1) that the traveling wave is stable when embedded into a finite lattice with 5 oscillators (minimal embedding) because all of its Floquet exponents have negative real parts. In its next higher embedding ($10$ oscillators), however, this traveling wave is unstable because it has a pair of Floquet exponents with positive real parts. Indeed, any higher embeddings beyond $5$ are also unstable because their Floquet exponents lie on a portion of the master stability curve that is on the positive real part of the half-plane. For $\alpha=0.8$ in panel (a2), the traveling wave is unstable even in its smallest embedding because there is a set of Floquet exponents with positive real part. Hence, there exists no embedding for which such a traveling wave is stable at this value of $\alpha$. This raises two natural questions: (a) when is the minimal embedding stable, and (b) does there exist a range of $q$ and $\alpha$ for which the traveling wave is stable across all embeddings.

\subsection{Stability of minimal embeddings and Eckhaus instability}
\label{sec:eckhaus}

The scenarios in panels (a1) and (a2) of Fig.~\ref{fig:eckhaus-curve} illustrate the influence of lattice size on the stability of traveling waves: even though segments of the master stability curve may extend into the positive half plane, the traveling wave corresponding to the minimal embedding of the profile may remain stable. 

Here, we will determine how the lattice size influences the stability and how to find parameters for which the master stability curve does not extend to the positive half-plane.
To determine when a given traveling wave is stable in smaller embeddings, we exploit the fact that the master stability curve is parameterized by a parameter $\phi$ such that the Floquet multipliers of the wave appear on the master stability curve in equal intervals of $\phi$~\cite{RuschelGiraldo}; see Sec.~\ref{sec:compEck} for more details. This allows us to impose the condition $\mathrm{Re}\, \lambda = 0$ when $\phi = 2\pi/N_{\min}$ ($N_{\min}$ is the minimum embedding size), so that we can compute the parameter values at which a traveling wave becomes unstable in its minimum embedding. In particular, we have found this instability to occur via a torus bifurcation. In this way, we can obtain the boundary of instability in the $(\alpha,q)$-parameter plane for different minimum lattice sizes. See Sec.~\ref{sec:compEck} for the details of the numerical implementation of this approach. 

Figure~\ref{fig:eckhaus-curve}(b1) displays the loci of torus bifurcations in lattices with $N_{\min}=N_1 = 5$ (cyan), $N_{\min}=N_2=10$ (darker cyan), and $N_{\min}=N_3=20$ (darkest cyan) oscillators in the $(\alpha,q)$ parameter plane. For a given number of oscillators $N$, there is a number of traveling waves with discrete wave numbers $q=2\pi M/N,$ where $M=1,2,\ldots,\lfloor N/2\rfloor$, which come into existence at a critical value of $\alpha>0$. The wave emerging at the first of these intersections (primary traveling wave) is stable immediately past the critical value and will remain stable for a certain range of $\alpha$, which decreases with increasing system size $N$. The cyan curves indicate the boundary of the regions in which the primary traveling wave is stable for all lattice sizes $N\leq N_1$ (cyan), $N\leq N_2$ (darker cyan), $N\leq N_3$ (darkest cyan). For this system, because these curves satisfy $\mathrm{Re},\lambda = 0$ for $\phi = 2\pi/N_{\min}$,  they also admit the following interpretation: primary traveling waves with wavenumber $q$ and parameter $\alpha$ located to the left of these curves (each associated with a given $N_{\min}$) are stable when embedded in lattices of size $N < N_{\min}$. Thus, these curves encode stability regions in the $(\alpha,q)$-parameter plane beyond a single wavenumber.

Notice in Fig.~\ref{fig:eckhaus-curve}~(b1) that, as $N_{\min}$ increases, the curves appear to accumulate along a well-defined fringe in the $(\alpha,q)$ parameter plane. This behavior is not fortuitous: as $N_{\min}$ increases, these curves approach the regions where the curvature of the master stability curve at $\lambda = 0$ changes sign. Indeed, we can extend the 2PBVP formulation to compute the limiting curve; see Sec.~\ref {sec:compEck} for more details on the numerical implementation.

In particular, the two blue curves in panel~(b1) indicate parameter values at which the curvature of the master stability curve changes sign. As shown in the zoomed view in panel~(b2), these curves delimit a region in which the master stability curve has strictly negative real part. This behavior is further illustrated in panels~(c1), (c2), and (c3), which display master stability curves near $\lambda = 0$ for the three parameter points marked in panel~(b2). Panel~(c2), corresponding to parameter values lying between the two blue curves, exhibits negative curvature near $\lambda = 0$; consequently, the associated traveling waves are stable for all admissible finite lattice embeddings. This observation has far-reaching consequences, as it implies that, in the presence of nonlinearity and nonreciprocity, there exist traveling waves that remain stable for arbitrarily large finite lattices.  From this perspective, the blue curves play a role analogous to the Eckhaus instability encountered in traveling waves of partial differential equations \cite{Rademacher2007,hasan2021spatiotemporal}, and destabilization of equilibria in lattices \cite {stableDiscrete}. We adopt the same naming convention here.

\begin{figure}
    \centering
    \includegraphics{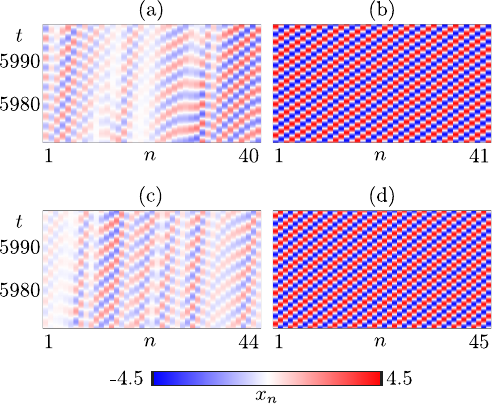}
    \caption{Spatiotemporal plots of traveling waves in system~(\ref{eq:main}) for $\alpha = 0.8$ in lattices with different size (a) N=40, (b) $N=41$, (c) $N=44$, and (d) $N=45$; all other parameters are as in Fig.~\ref{fig:eckhaus-curve}. The initial conditions are chosen close to the existing traveling wave solutions (which are unstable in (a) and (c)).} \label{fig:examplePerSol} 
\end{figure}

To demonstrate the versatility of using Eckhaus instability curves in discrete systems, we illustrate their application in determining the influence of lattice size on wave stability. For $\alpha = 0.8$, Fig.~\ref{fig:eckhaus-curve}(b2) shows that the interval of wave numbers for which the traveling wave is stable is approximately $$0.2437 \leq q/2\pi\leq 0.2455.$$
From this interval, we can compute the minimum lattice size $N_d$ such that any lattice of size greater than or equal to $N_d$ can support a stable traveling wave, namely
$$N_d=\left\lfloor \frac{1}{0.2455-0.2437} \right\rfloor+1=556.$$
In particular, for $N \geq N_d$, one can always find wave numbers of the form $q = 2\pi k = 2\pi M/N$ that fall within the stability interval, i.e.,$$k=q/2\pi \in (0.2437 ,0.2455).$$ 
Heuristically, the numerator $M$ corresponds to the number of repetitions of the fundamental periodic profile composing the traveling wave; for example, if the periodic profile resembled a pulse, then $M$ would represent the number of pulses forming the wave.

It is important to note that smaller lattice sizes may still admit stable traveling waves with wave numbers in the interval $0.2437 \leq q/2\pi \leq 0.2455$. For instance, lattices of sizes $N = 41$ and $N = 45$ support traveling waves with wave numbers $q/2\pi = 10/41 \approx 0.2439$ and $q/2\pi = 11/45 \approx 0.2444$, respectively. In these cases, the numerators $10$ and $11$ indicate the number of identical wave profiles propagating in the finite lattice.

We refer to lattice sizes below $N_d$ for which traveling waves remain stable as sporadic sizes. Figure~\ref{fig:examplePerSol} presents the results of numerical simulations that illustrate this phenomenon: stable traveling waves propagate in lattices of sizes $N = 41$ and $N = 45$, whereas they are unstable for $N = 40$ and $N = 44$, with nearby solutions eventually diverging. These simulations were obtained by initializing the system close to the traveling wave solution and applying a small random perturbation. To identify the sporadic sizes associated with a given interval of wave numbers $q/2\pi \in (q_{\min}, q_{\max})$, we proceed as follows:
\begin{enumerate}
\item Compute $N_d$ as described above.
\item For each $N < N_d$, compute $\lceil N q_{\min} \rceil$ and $\lfloor N q_{\max} \rfloor$.
\item If $\lceil N q_{\min} \rceil \leq \lfloor N q_{\max} \rfloor$, then $N$ is a sporadic size.
\end{enumerate}

\begin{figure}
    \centering
    \includegraphics{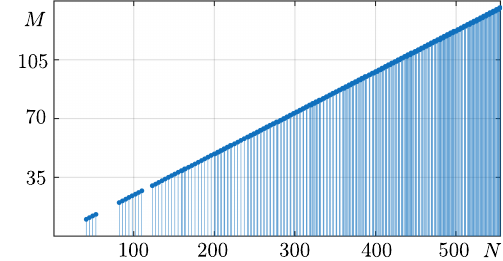}
    \caption{Sporadic stable lattice sizes $N$ for a fixed value of $\alpha=0.8$, displayed via the numerator index $M$, i.e, $0.2437\leq2\pi M/N\leq 0.2455$. Each point corresponds to a lattice size admitting a stable traveling wave. The irregular distribution reflects the non-monotonic dependence of stability on $N$.} \label{fig:sporadic} 
\end{figure}

Figure~\ref{fig:sporadic} shows the sporadic stable sizes obtained by applying the previous algorithm at $\alpha = 0.8$ over the interval $0.2437 \leq q/(2\pi) \leq 0.2455$. For $N>566$, it is possible to obtain more than one embedding dimension $MN,(M+1)N,\ldots$ for a given $N$ so that the waves with wavenumber $M/N,(M+1)/N,...$ are stable. 

In summary, the Eckhaus instability framework makes it possible to carry out quantitative analysis of size effects, specifically the bounds on lattice sizes for which a member of a given family of traveling waves is stable. 

\section{Quasiperiodic traveling waves}
\label{sec:anharmonic}

It is common to observe quasiperiodic traveling waves in coupled nonlinear oscillators~\cite{coulaisOddNonlinear,playground}. Bifurcation analysis of quasiperiodic waves is much more involved than the analysis of time-periodic waves. To use the methodology described in Section~\ref{sec:computation} for the analysis of quasiperiodic traveling waves, we derive the equations that govern the envelope of the response. This is helpful because the envelope of a quasiperiodic traveling wave may be periodic in time; see also~\cite{BehroozPRB}. In particular, the envelope equations can be used to determine the stability of the quasiperiodic traveling waves.

\subsection{Rotating wave approximation}

We consider an approximation to periodic and quasi-periodic traveling waves exhibited by system~\eqref{eq:main}, under which solutions of system~\eqref{eq:main}  can be represented as twice the real part of a complex rotating wave $A_n(t)e^{i\omega_0 t}$. That is, we consider the ansatz 
\begin{equation}\label{eq:anzFrequency}
x_n(t) = A_n(t)e^{i\omega_0 t}+ \overline{A_n(t)}e^{-i \omega_0 t},
\end{equation}
where $\omega_0$ can be thought of as the frequency of the periodic oscillations at a given location $n$. In this approximation, we choose $\omega_0$ such that it captures the fast frequency of the oscillations, while any periodicity in the coefficient functions $A_n(t)$ captures the dynamics of the envelope, which are slow compared to $\omega_0$.  


Assuming that the (non-resonant) terms of frequency $3\omega_0$ are negligible, and separating the complex conjugate terms, the rotating wave ansatz results in the following equation governing the dynamics of the response envelope, as characterized by the coefficient functions $A_n$
\begin{equation}
\label{eq:envMainFreq}
\begin{aligned}
0=&\ddot{A}_n + 2i\omega_0\dot{A}_n +2\zeta\dot{A}_n+ 2i\omega_0\zeta A_n+(1-\omega_0^2)A_n \\
&+\kappa (2A_n-A_{n+1}-A_{n-1})\\
&+\alpha \kappa (A_{n+1}-A_{n-1}) +3\beta |A_n|^2A_n,
\end{aligned}
\end{equation}
We refer to this as the \emph{rotating wave approximation}, RWA. In this work, we will consider system~\eqref{eq:envMainFreq} as an envelope equation, that is, we expect $2|A_n|$ to capture the envelope of the $x$-oscillations. 

We note that system~\eqref{eq:envMainFreq} can be brought into the form of a complex nonlinear Schrödinger equation by further approximation steps, namely, by neglecting the second derivative term $\ddot{A}_n$ and the detuning term $(1-\omega_0^2)A_n$. These additional approximations further reduce the accuracy of the equation in capturing the correct envelope of the response, specifically limiting in the case of lattices of small and moderate sizes.
Notice that this improved accuracy in capturing the response envelope comes at the expense of increasing the dimension of the system from $2N$ degrees of freedom in system~\eqref{eq:main} to $4N$ in system~\eqref{eq:envMainFreq}.
Furthermore,  in order to have a good approximation of a periodic traveling waves in system~\eqref{eq:main}, we need to identify a particular $\omega_0$. This can be achieved, e.g. by estimating $\omega_0$ from simulation such that the phase of each $A_n$. 

In what follows, we specifically refer to the RWA to avoid confusion with nonlinear Schrödinger-type approximations typically considered as envelope equations.

For the analysis in this section, it is convenient to consider system~\eqref{eq:envMainFreq} in polar coordinates. To do so, we define
$$
A_n(t)=R_n(t)e^{i\phi_n(t)}.
$$
with $R_n(t)\ge 0, \phi_n(t)\in\mathbb{R}$. After algebraic manipulation, we obtain the RWA in polar coordinates:
\begin{align}
0={}&\ddot R_n - R_n\dot\phi_n^2
-2\omega_0 R_n\dot\phi_n
+2\zeta\dot R_n
+(1-\omega_0^2)R_n
\nonumber\\
&+\kappa\Bigl(
2R_n
-R_{n+1}\cos(\Delta_n^+)
-R_{n-1}\cos(\Delta_n^-)
\Bigr)
\nonumber\\
&+\alpha\kappa\Bigl(
R_{n+1}\cos(\Delta_n^+)
-R_{n-1}\cos(\Delta_n^-)
\Bigr)
+3\beta R_n^3,
\label{eq:polar_real_secondorder}
\\[0.6em]
0={}&2\dot R_n\dot\phi_n + R_n\ddot\phi_n
+2\omega_0\dot R_n
+2\zeta R_n\dot\phi_n
+2\omega_0\zeta R_n
\nonumber\\
&+\kappa\Bigl(
-R_{n+1}\sin(\Delta_n^+)
-R_{n-1}\sin(\Delta_n^-)
\Bigr)
\nonumber\\
&+\alpha\kappa\Bigl(
R_{n+1}\sin(\Delta_n^+)
-R_{n-1}\sin(\Delta_n^-)
\Bigr).
\label{eq:polar_imag_secondorder}
\end{align}
where $
\Delta_n^+ := \phi_{n+1}-\phi_n$ and $\Delta_n^- := \phi_{n-1}-\phi_n$ are introduced for brevity. 

\subsection{Existence of rotating waves}

We consider a rotating wave of the form $A_n = A e^{iqn}$ as a relative equilibrium in the RWA. Substituting this into system~\eqref{eq:envMainFreq} and separating the real and imaginary parts gives the following dispersion relation
$$
\begin{aligned}
(1-\omega_0^2) + \kappa(2-2\cos q) + 3\beta r^2 = 0, \\
2\zeta\omega_0 + 2\alpha\kappa \sin q = 0,
\end{aligned}
$$
From the real part, the wave amplitude as a function of the wavenumber is
$$
r^2 = \frac{1}{3\beta}\left(\omega_0(q)^2 - 1 - 2\kappa(1-\cos q)\right),
$$
with the constraint that the right-hand side is positive.
From the imaginary part, the frequency can be expressed as a function of the wavenumber $q$ as
$$
\omega_0(q) = -\frac{\alpha\kappa}{\zeta} \sin q, \qquad (\zeta \neq 0).
$$

As in Section~III.B, we see that a balance between $\alpha$ and $\zeta$ is important for traveling waves to exist. If $\zeta=0$, then the condition reduces to $\alpha\kappa \sin q = 0$, requiring either $\alpha\kappa=0$ or $q \in \{0,\pi\}$.

For a finite ring with $N$ units, the periodic boundary conditions $A_{n+N}=A_n$ restrict the allowed wavenumbers to
$$
q = \frac{2\pi m}{N}, \qquad m=0,1,\dots,N-1.
$$
The corresponding frequencies and amplitudes associated with these wavenumbers represent the existence conditions for RWs of the envelope equation of the lattice. These RWs are non-zero states of constant amplitude and different relative phases. In the original system~\eqref{eq:main}, they represent periodic traveling waves; periodic solutions at each lattice site with different phases. 

\subsection{Spectrum and stability of rotating waves}

The stability of RWs can be analyzed by considering the linear behavior of small perturbations applied to a RW. Thus, we consider $A_n(t)=R_n(t) e^{iqn}$ with
$R_n(t)=r+\rho_n(t)$ and $ \phi_n(t)=qn+\theta_n(t)
$; see also~\cite{MI_KivsharPeyrard,burlakov1998,dauxoisKhomerikiRuffo}.
Using the RWA in the polar coordinate, Eqs.~\eqref{eq:polar_real_secondorder}--\eqref{eq:polar_imag_secondorder}, we obtain the following linear equations governing the perturbations:
\begin{align}
0={}&\ddot\rho_n + 2\zeta\dot\rho_n - 2\omega_0 r\,\dot\theta_n
+\Bigl[(1-\omega_0^2)+2\kappa+9\beta r^2\Bigr]\rho_n
\nonumber\\
&-\kappa\cos q\,(\rho_{n+1}+\rho_{n-1})
+\alpha\kappa\cos q\,(\rho_{n+1}-\rho_{n-1})
\nonumber\\
&+\kappa r\sin q\Bigl[(\theta_{n+1}-\theta_n)-(\theta_{n-1}-\theta_n)\Bigr]\nonumber\\
&-\alpha\kappa r\sin q\Bigl[(\theta_{n+1}-\theta_n)+(\theta_{n-1}-\theta_n)\Bigr],
\label{eq:lin_real_secondorder}
\end{align}
\begin{align}
0={}&r\,\ddot\theta_n + 2\omega_0\dot\rho_n + 2\zeta r\,\dot\theta_n + 2\omega_0\zeta\,\rho_n
\nonumber\\
&+\kappa\sin q\,(-\rho_{n+1}+\rho_{n-1})
+\alpha\kappa\sin q\,(\rho_{n+1}+\rho_{n-1})
\nonumber\\
&-\kappa r\cos q\Bigl[(\theta_{n+1}-\theta_n)+(\theta_{n-1}-\theta_n)\Bigr]\nonumber\\
&+\alpha\kappa r\cos q\Bigl[(\theta_{n+1}-\theta_n)-(\theta_{n-1}-\theta_n)\Bigr].
\label{eq:lin_imag_secondorder}
\end{align}

As system \eqref{eq:lin_real_secondorder}-\eqref{eq:lin_imag_secondorder} is linear and invariant under a cyclic index shift (periodic boundary conditions), the perturbations are of the form
$$
\rho_n(t)=\rho\,e^{\lambda t+ipn},\qquad
\theta_n(t)=\theta\,e^{\lambda t+ipn},$$
with growth rate $\lambda\in\mathbb{C}$ and perturbation wavenumber
$\qquad p\in[-\pi,\pi]$. We can therefore obtain the following linear system for the perturbations of a RW:
$$
M(\lambda,p)\begin{pmatrix}\rho\\ \theta\end{pmatrix}=0,
$$
with
$$
M(\lambda,p)=
\begin{pmatrix}
m_{11}(\lambda,p) & m_{12}(\lambda,p)\\
m_{21}(\lambda,p) & m_{22}(\lambda,p)
\end{pmatrix},
$$
where
\begin{align}
m_{11}(\lambda,p)
=&\lambda^2+2\zeta\lambda+(1-\omega_0^2)+2\kappa+9\beta r^2 \nonumber\\
&-2\kappa\cos q\,\cos p 
+2i\alpha\kappa\cos q\,\sin p, 
\label{eq:m11_secondorder}
\\[0.3em]
m_{12}(\lambda,p)
=&-2\omega_0 r\,\lambda
+2i\kappa r\sin q\,\sin p \nonumber\\
&-2\alpha\kappa r\sin q\,(\cos p-1),
\label{eq:m12_secondorder}
\\[0.3em]
m_{21}(\lambda,p)
=&2\omega_0\lambda+2\omega_0\zeta
-2i\kappa\sin q\,\sin p \nonumber\\
&+2\alpha\kappa\sin q\,\cos p,
\label{eq:m21_secondorder}
\\[0.3em]
m_{22}(\lambda,p)
=&r\lambda^2+2\zeta r\,\lambda
-2\kappa r\cos q\,(\cos p-1) \nonumber\\
& +2i\alpha\kappa r\cos q\,\sin p.
\label{eq:m22_secondorder}
\end{align}
Nontrivial solutions to the variational problem \eqref{eq:lin_real_secondorder}--\eqref{eq:lin_imag_secondorder} exist if and only if
\begin{equation}
\label{eq:dispersion_det_secondorder}
\begin{aligned}
0=&\det M(\lambda,p)\\
=&m_{11}(\lambda,p)m_{22}(\lambda,p)-m_{12}(\lambda,p)m_{21}(\lambda,p),
\end{aligned}
\end{equation}
for a given $p=2\pi m/N,$ $m=1,\ldots,N$. Eq.~\eqref{eq:dispersion_det_secondorder} defines a relation between the growth rate and the wavenumber of the perturbations to the rotating waves. It also describes curves $\lambda=\lambda(p)$ when considering $p$ as a continuous parameter, $p\in[0,2\pi]$. In particular, when the real part of $\lambda$ is zero, these curves correspond to the onset of instability of the rotating waves.  

To stress the analogy with the Floquet spectrum of traveling waves considered earlier, we also refer to the curves described by $\lambda=\lambda(p)$ (Floquet-Bloch spectrum) as master stability curves. The master stability curves provide a direct characterization of the spectral response of a given rotating wave solution with wave number $q$ to a plane-wave perturbation of (angular) wave number $p$.

\begin{figure}
    \centering
    \includegraphics[width=\linewidth]{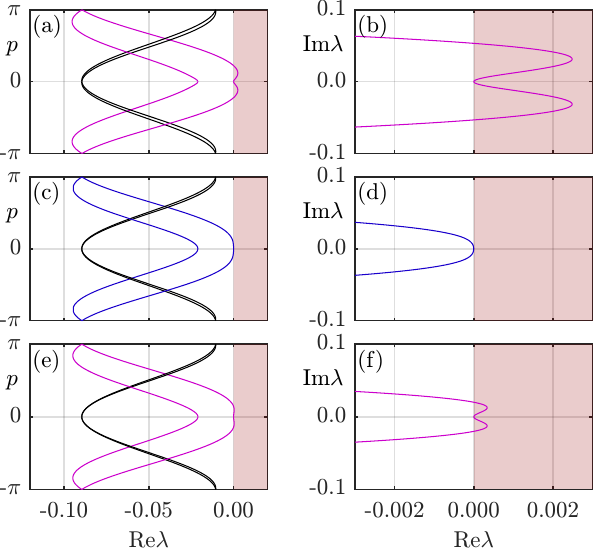}
    \caption{Master stability curves (magenta, blue, black) of rotating waves at $\alpha=0.8$, (a)--(b) $q=0.24$ (unstable), (c)--(d) $q=0.2445$ (stable), and (c)--(d) $q=0.244$ (unstable); cf. Fig. \ref{fig:eckhaus-curve}(c). Panels (a),(c),(e) show, in projection, the respective growth rate of a given (angular) perturbation wave number $p$ of the rotating wave. Panels (b),(c),(f) show an enlargement of the spectrum (projection onto real and imaginary part) close to zero. Other parameter are $\beta  = 0.1$, $\zeta  = 0.05$, $\kappa = 0.1$.}  \label{fig:RW-spectrum}
\end{figure}

In Fig.~\ref{fig:RW-spectrum}, we plot the master stability curves for the parameter values corresponding to Fig.~\ref{fig:eckhaus-curve}(c) for three different wave numbers (a)--(b) $q = 0.24$, (c)--(d) $q = 0.2445$, and (e)--(f) $q = 0.244$. For $q = 0.24$, the spectrum exhibits eigenvalues with positive real parts over a finite band of wave numbers $p$, indicating linear instability of the rotating wave for sufficiently large $N$. As $q$ increases to $q = 0.2445$, the master stability curve attached to $\lambda=0$ locally changes its curvature. As a result, $\mathrm{Re}\,\lambda(p) < 0$ for all admissible $p$ and the corresponding RW becomes spectrally stable irrespective of the lattice size $N$. Further decreasing the RW wave number to $q = 0.244$ causes the curvature to change once more, which reintroduces eigenvalues with small positive real parts for a sufficiently large lattice size. 

Note that Figs.~\ref{fig:RW-spectrum} (b), (d) and (e) show the same qualitative change as the periodic traveling wave solution in system~\ref{eq:main} for the chosen parameter values; cf. Fig.~\ref{fig:eckhaus-curve}(c). This strongly suggests that the RW solutions of system~\eqref{eq:envMainFreq} exhibit an Eckhaus phenomenon; and, conversely, that the Eckhaus phenomenon in system~\eqref{eq:main} is well captured by system~\eqref{eq:envMainFreq}. Motivated by this observation, we obtain analytical expressions for the corresponding Eckhaus curves and compare the predictions to the Eckhaus curve computed in Fig.~\ref{fig:eckhaus-curve}(b2).

To find the locus of the Eckhaus instability, we look for points of zero curvature of the master stability curve at $\lambda=0$. 
Equation~(\ref{eq:dispersion_det_secondorder}) defines a function $f(\lambda(p),p)$, where $\lambda$ depends on $p$. By implicit differentiation, we obtain
$$\lambda_{pp}=-\left.\frac{f_{\lambda\lambda} \cdot (f_p/f_\lambda)^2+2f_{\lambda p}(f_p/f_\lambda) +f_{pp}}{f_\lambda}\right|_{(p,\lambda)=(0,0)}.$$
which we rewrite as
\begin{equation}
\lambda_{pp} = p_1p_2,
\end{equation}
where 
$$p_1=-\frac{\kappa \bigl(\zeta^2 - \alpha^2 \kappa \cos q\bigr)}{
\sin^2 q\, \alpha^2 \zeta^3 \kappa^2
+ \zeta^5 \bigl(-1 + 2(-1+\cos q)\kappa\bigr)}$$
and 
\begin{align*}
    p_2=&\Bigl(- \zeta^2 \cos q + 2\zeta^2 \kappa (({-1+\cos q})\cos q - \sin^2 q)+  \\
    & \quad 3\alpha^2 \kappa^2 \cos q\, \sin^2 q \Bigr).
\end{align*}
Looking for $\lambda_{pp}=0$, we find that the condition $p_2=0$ defines the analytical Eckhaus curves.

\begin{figure}[!]
    \centering
    \includegraphics{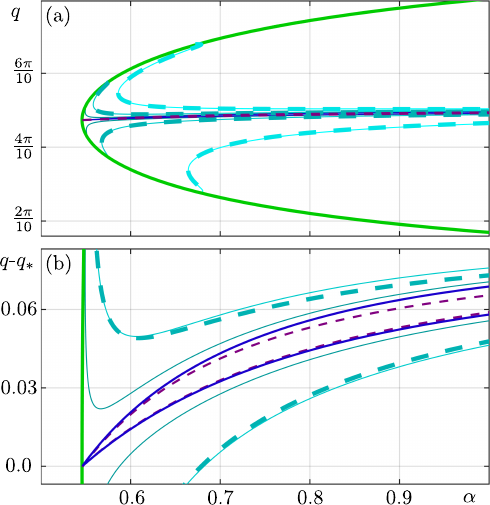}
    \caption{Panel~(a) shows the bifurcation diagram in the $(\alpha,q)$ parameter plane, as presented in Fig.~\ref{fig:eckhaus-curve}(b1). The dashed lines correspond to analytical expressions of the bifurcation curves computed from \eqref{eq:dispersion_det_secondorder}; their colors match those of the curves obtained from the Duffing equation. Panel~(b) presents an enlargement of the Eckhaus region, where the dashed lines again represent the analytical expressions derived from \eqref{eq:dispersion_det_secondorder}.}
    \label{fig:AnalyticalComparison}
\end{figure}

We can also use use Eq.~\eqref{eq:dispersion_det_secondorder} to find the loci of torus bifurcation points. To do so, we employ an approach from computational algebraic geometry~\cite{NAG}. 
Considering the function $f(\lambda(p),p)$, we introduce the substitutions $a:=\cos q$ and $b:=\sin q$, and evaluate the function at $\lambda(p)=i v$. We fix $p=\tfrac{2\pi}{5}$ or $p=\tfrac{2\pi}{10}$, corresponding to $N_{\min}=5$ or $10$ respectively. We recast $f$ as a polynomial function $h(a,b,v)$ and compute a Gröbner basis for the system
\[
\{\operatorname{Re} h(a,b,v),\ \operatorname{Im} h(a,b,v),\ a^2 + b^2 - 1\} = \{0,0,0\},
\]
under lexicographical ordering $v > \alpha > a > b$. The second element of this basis (the first one is $a^2 + b^2 - 1$)  yields an implicit relation between $a, b$ and $\alpha$, which we use to obtain the dashed curves in Fig.~\ref{fig:AnalyticalComparison}; see the Mathematica notebook in the provided repository.

Fig.~\ref{fig:AnalyticalComparison} compares the bifurcation diagrams predicted by the preceding analysis to the ones computed by numerical continuation. The solid curves are from Fig.~\ref{fig:eckhaus-curve}, which include the Hopf curve (green), loci of torus bifurcations (cyan) and the Eckhaus instability curve (blue). The dashed curves represent the analytical predictions of the same loci derived from \eqref{eq:dispersion_det_secondorder}. The good agreement between the computational (based on continuation) and analytical (based on RWA) results indicates that the RWA accurately captures the dynamics of the traveling waves. 

\subsection{Stability analysis of modulated waves}

Another key advantage of system~\eqref{eq:envMainFreq} is that it enables numerical continuation techniques \cite{KrauskopfOsingaBook2007} to track the modulated wave, i.e., the envelope of the traveling wave. This allows us to characterize how the envelope evolves as the wavenumber $q$ varies, and to explain the behavior observed in Fig.~\ref{fig:1}~(e) and (g).

\begin{figure}[!]
    \centering
    \includegraphics{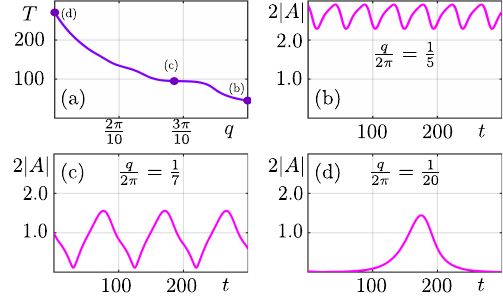}
    \caption{Period and amplitude of modulated waves in system~\eqref{eq:envMainFreq} for $\alpha = 0.75$. Panel~(a) shows the period of the modulated waves as the wavenumber decreases. Panels~(b)--(d) display twice the amplitude of the corresponding modulated waves at the wavenumbers indicated in each panel.}
    \label{fig:profilesRWA}
\end{figure}

Figure~\ref{fig:profilesRWA}(a) shows how the period $T$ of the rotating wave (traveling wave envelope) depends on the wavenumber for $\alpha = 0.75$. In particular, the period increases as $q$ decreases. Panels~(b)–(d) illustrate how the profile of the rotating wave becomes increasingly localized as $q$ decreases. Equivalently, as the lattice size $N$ increases, it can support waves with increasingly smaller wavenumbers and localization emerges. This behavior is also observed in the transient dynamics shown in Fig.~\ref{fig:1}(g1), where the trajectory localized for a period of time.

The envelopes such as those in Fig.~\ref{fig:profilesRWA} may be either stable or unstable. For example, the envelope is stable in Fig.~\ref{fig:1}(f), whereas it is unstable in Fig.~\ref{fig:1}(g). Because rotating waves are traveling-wave solutions of system~\eqref{eq:envMainFreq}, their stability can be readily assessed within the master stability framework.

\begin{figure}[!]
    \centering
    \includegraphics{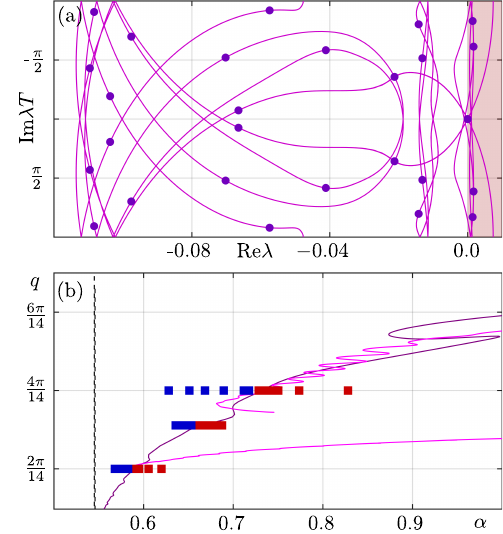}
    \caption{Panel~(a) shows the Floquet spectrum (purple dots) for an embedding in a lattice of size $7$, and master stability curves (magenta), of the modulated wave shown in Fig.~\ref{fig:profilesRWA}(c); here $(\alpha,q)=(0.75,2\pi/7)$. Panel (b) shows two-parameter continuation in the $(\alpha,q)$-plane of torus bifurcation curves (purple, magenta) corresponding to the crossing of leading unstable Floquet exponents shown in panel (a); also shown are the locus of representative stable (unstable) modulated waves indicated by blue (red) squares obtained by sampling along branches of modulated waves using numerical continuation. The critical value $\alpha_\ast$ is indicated by a black dashed curve.}   \label{fig:RWAStability}
\end{figure}

Figure~\ref{fig:RWAStability}~(a) shows the Floquet spectrum and the corresponding master stability curve for the rotating wave with $q = 2\pi/7$ and $\alpha = 0.75$ in a lattice of size $N = 7$; the wave in Fig.~\ref{fig:profilesRWA}~(c). There are four Floquet exponents with positive real parts, indicating instability. More generally, we find that a portion of the master stability curves lies in the positive half-plane for different values of $\alpha$ and $q$, implying that embeddings in larger lattices are necessarily unstable in the case of rotating waves.

For smaller values of $\alpha$, however, the minimal embedding of the rotating wave may be stable. Analogous to the computation of the loci of torus bifurcation points in Fig.~\ref{fig:eckhaus-curve}~(b1), we determine the loci at which the minimal embedding loses stability through a torus bifurcation. For system~\eqref{eq:envMainFreq}, this analysis is more involved (compared to system~(\ref{eq:main})) because multiple curves must be tracked and their intersections with the imaginary axis identified. 

For the rotating wave shown in Figure~\ref{fig:RWAStability}~(a), there are two branches of the Floquet spectrum that cross the imaginary axis. 
Figure~\ref{fig:RWAStability}(b) shows the two torus bifurcation curves (purple and magenta) that mark the parameter values at which the Floquet exponents of these two branches become purely imaginary. 
Any higher value of $\alpha$ yields an unstable modulated wave in any embedding dimension. Conversely, to left of the torus curves, modulated waves with wave numbers $k=1/N$ might be stable when embedded in a lattice with $N$ nodes, at least in the vicinity of the bifurcation point. 
We further validate this claim by taking representative points at three different wavenumbers $q/(2\pi)=1/7,1/9,1/14$ and checking their stability; these are shown by the blue (stable) and red (unstable) square markers. The stable modulated wave indicated at $(\alpha,q)=(0.61,2\pi/14)$, is shown in Fig.~\ref{fig:1}(f1),(f3), in comparison with the corresponding underlying quasiperiodic traveling wave Fig.~\ref{fig:1}(f1-f2).

We have observed that additional transition curves may reappear at lower wavenumbers, corresponding to other branches of the Floquet spectrum crossing the imaginary axis. We have included one such transition curve in Figure~\ref{fig:RWAStability}(b); starting near $q=2\pi/14$. For illustration purposes, we have only included three transition curves, but we remark that there exist more. 

\section{Master stability for traveling waves}\label{sec:compEck}

It has recently been shown that the spectral stability of traveling waves on rings and lattices can be computed efficiently via numerical continuation \cite{RuschelGiraldo}. The key idea is to exploit the translational symmetry of the lattice to block–diagonalize the variational problem around a traveling wave using a spatial Fourier transform. In this formulation, the perturbation wave number appears as an additional spectral parameter, which can be considered as continuous, allowing the stability problem to be treated by continuation rather than by constructing the full monodromy matrix. The approach can therefore be viewed as a Floquet–Bloch decomposition of perturbations around a nonlinear traveling wave, extending the classical Floquet–Bloch theory for ground states on periodic and periodically driven lattices to traveling waves \cite{gomez2013floquet, porter2005nonlinear}. In this section, we will briefly summarize the master stability framework and showcase how one can extend it to compute the boundaries of Eckhaus instability for system~\eqref{eq:main} and system~\eqref{eq:envMainFreq}.

\subsection{Master stability curve}
A traveling wave solution of system~\eqref{eq:main} is a periodic solution with spatiotemporal symmetry in which each oscillator follows the same wave profile but shifted in time along the lattice. Such solutions can be written as $x_n(t)=\tilde x(t-n\tau),$
where $\tilde x(t)$ is the wave profile, $T$ is its period, and $\tau$ denotes the phase delay between neighboring oscillators. The associated wave number is therefore
$$
k=\frac{\tau}{T}.
$$
Substituting this ansatz into system~\eqref{eq:main} reduces the lattice dynamics to the single advanced-delayed equation for the profile $\tilde x(t)$:
\begin{equation}\label{eq:travelingWaveAnsatz}
\begin{aligned}
    \ddot{\tilde x}(t) + 2\zeta\dot{\tilde x}(t)+ \tilde{x}(t) + \kappa (2\tilde{x}(t)-\tilde{x}(t-\tau)-\tilde{x}(t+\tau)) \\
    + \alpha \kappa (\tilde{x}(t-\tau)-\tilde{x}(t+\tau)) 
    + \beta \tilde{x}^3(t)= 0,
\end{aligned}
\end{equation}
which we represent as a first-order system in the variable $\tilde z=(\tilde x,\dot{\tilde{x}})$ as
$$\dot{\tilde{z}} = G(\tilde{z}(t+\tau),\tilde{z}(t),\tilde{z}(t-\tau)).$$

To determine the stability of the traveling wave, one considers its  Floquet exponents \cite{kuznetsov}. As shown in \cite{RuschelGiraldo}, the translational invariance of system~\eqref{eq:main} under lattice shifts allows one to compute each Floquet exponent $\lambda$ with respect to a spatial perturbation wave number (spatial Fourier modes of the lattice). Following \cite{RuschelGiraldo}, the stability can be assessed by determining the values of $\lambda$ for which
$$
\zeta'(t)=-\lambda\zeta(t)+\sum_{-1\leq m\leq 1} e^{im\phi}B_{m}(t)\zeta(t-m\tau),
$$
admits $T$-periodic solutions, i.e., $\zeta(t+T)=\zeta(t)$ for a given $\phi\in[0,2\pi)$; where $B_m(t)=D_{m+2}G(\tilde{z}(t+\tau),\tilde{z}(t),\tilde{z}(t-\tau))$ and $D_{l}G$ represents the partial derivatives matrix of $G$ with respect to its $l$-th two-dimensional argument. We remark that the $B_m(t)$ are matrices with $T$-periodic components, i.e., $B_m(t)=B_m(t+T)$.

In this setting, each pair $(\lambda,\zeta)$ ---consisting of a Floquet exponent and its associated periodic solution--- depends on the parameter $\phi$, which represents the spatial phase (perturbation wave number). 
For each $\phi$, this equation yields Floquet exponents $\lambda(\phi)$ describing the growth or decay of perturbations with perturbation wave number $\phi$. The entire stability spectrum of the traveling wave can therefore be represented by
$$
\lambda=\lambda(\phi),
$$
which we refer to as the master stability curve \cite{RuschelGiraldo}. For a lattice with $N$ oscillators, finitely many values $\phi=2\pi l/N$, $l=0,\dots,N-1$, give rise to the full Floquet spectrum of the wave in the finite lattice problem. The master stability curve, therefore, acts as a continuous envelope of the Floquet spectrum of waves with wave number $k$ for lattices of arbitrary size. Instability occurs when $\lambda(\phi)$ crosses the imaginary axis with positive speed, implying $\mathrm{Re}\,\lambda(\phi)>0$ for some $\phi$, corresponding to the growth of spatially modulated perturbations. Examples of such instabilities include the Turing instability and the Eckhaus scenario (long-range/sideband instabilities) \cite{cross1993pattern}.

In practice, the master stability curve is computed via numerical continuation of a suitable two–point boundary value problem (2PBVP) \cite{KrauskopfOsingaBook2007}, using the perturbation wave number $\phi$ as the continuation parameter. This approach avoids the explicit construction of the lattice monodromy matrix and reduces the stability analysis of system~\eqref{eq:main} to a low-dimensional periodic boundary–value problem whose size depends only on the oscillator dynamics, rather than on the number of lattice sites. Specifically, one formulates a coupled boundary–value problem for the traveling wave profile $\tilde z = (\tilde x, \dot{\tilde{x}})$ and the perturbation mode $\zeta$, together with appropriate boundary, phase and normalization zero conditions. Accordingly, the 2PBVP takes the form
\begin{align}
    \tilde{z}^\prime(t) =& TG\left(\tilde{z}\left(t+\frac{\tau}{T}\right),\tilde{z}(t),\tilde{z}\left(t-\frac{\tau}{T}\right)\right), \label{eq:BVP1}\\
    \zeta^\prime(t) =& T\!\left(\!-\lambda \zeta(t) + \!\!\!\!\!\!\!\sum_{-1\leq m \leq 1 } \!\!\!\!\!  e^{im\phi} B_m(t)\zeta\left(t-m\frac{\tau}{T}\right)\!\right), \label{eq:BVP2}\\
    \tilde{z}(0) =& \tilde{z}(1), \label{eq:BVP3}\\
    \zeta(0) =& \zeta(1), \label{eq:BVP4}\\
    0 =& \int_0^1 \langle \tilde{z}, \tilde{z}'_{\text{ref}}\rangle + \langle \zeta, \zeta'_{\text{ref}}\rangle\,dt, \label{eq:BVP5}\\
    1 =& T||\zeta||_{L_2}^2, \label{eq:BVP6}\\
    0 =& \text{Im}(\pi_1(\zeta(0))), \label{eq:BVP7} \\
    0 =& \tau - k T. \label{eq:BVP8}
\end{align}

Equations (\ref{eq:BVP1})–(\ref{eq:BVP8}) define a well-posed two–point boundary value problem for the traveling wave profile and its perturbation, with the period $T$ appearing explicitly as a parameter and periodic boundary conditions (\ref{eq:BVP3})–(\ref{eq:BVP4}) imposed on the rescaled interval $[0,1]$. The formulation includes a phase condition (\ref{eq:BVP5}) to remove time-translation invariance, as well as normalization constraints that fix the scaling (\ref{eq:BVP6}) and phase (\ref{eq:BVP7}) of the perturbation mode. Condition~(\ref{eq:BVP8}) is added to monitor and keep track of the wave number of the traveling wave. In the 2PBVP, $\tilde z_{\mathrm{ref}}$ and $\zeta_{\mathrm{ref}}$ denote reference solutions from the previous continuation step, and $\pi_1$ denotes the projection onto the first component.

By allowing $(\tilde z,\zeta,\phi,\lambda,\tau,T)$ to vary under a pseudo-arclength continuation procedure, one computes the master stability curve; see \cite{RuschelGiraldo} for implementation details and \cite{KrauskopfOsingaBook2007} for an introduction to continuation methods. In practice, we use the software package \textsc{DDE-Biftool} for Matlab/Octave \cite{Sieber2014}, which provides discretization and continuation routines for advanced-delayed differential equations.

Using the 2PBVP defined in (\ref{eq:BVP1})–(\ref{eq:BVP8}), we compute the master stability curves shown in Fig.~\ref{fig:eckhaus-curve}. As noted above, the rotating wave approximation admits inhomogeneous traveling waves (modulated waves) of the form  $A_n(t)=\tilde A(t-n\tau)e^{-2\pi i n k}$, under which system~(\ref{eq:envMainFreq}) reduces to the advanced-delayed equation
\begin{equation}\label{eq:envMainFreq2}
\begin{aligned}
0=&\ddot{\tilde{A}}(t) + 2i\omega_0\dot{\tilde{A}}(t) +2\zeta\dot{\tilde{A}}(t)+ 2i\omega_0\zeta \tilde{A}(t)\\
&+(1-\omega_0^2)\tilde{A}(t)+3\beta |\tilde{A}(t)|^2\tilde{A}(t)\\
& +\kappa (2\tilde{A}(t)-e^{-2\pi i k}\tilde{A}(t-\tau)-e^{2\pi i k}\tilde{A}(t+\tau))\\
&+\alpha \kappa (e^{-2\pi i k}\tilde{A}(t-\tau)-e^{2\pi i k}\tilde{A}(t+\tau)).
\end{aligned}
\end{equation}

Introducing $\tilde{z}=(\text{Re}(\tilde{A}),\text{Im}(\tilde{A}),\text{Re}(\dot{\tilde{A}}),\text{Im}(\dot{\tilde{A}}))$, system~\eqref{eq:envMainFreq2} can again be written as the first-order system
$$\dot{\tilde{z}} = G(\tilde{z}(t+\tau),\tilde{z}(t),\tilde{z}(t-\tau)).$$
Thus, the 2PBVP (\ref{eq:BVP1})–(\ref{eq:BVP8}) can be used to compute the corresponding master stability curves. However, due to the additional $S^1$-equivariance of system~(\ref{eq:envMainFreq2})—namely, $\tilde A \mapsto e^{i\theta}\tilde A$—an extra phase condition is required to obtain a well-posed problem. We therefore impose
\begin{equation}\label{eq:BVP9}
\int_0^T \pi_2(z) dt =0
\end{equation}
which fixes the phase by enforcing a zero integral of the imaginary component. With this additional constraint, the master stability curve is computed by continuing $(\tilde z,\zeta,\phi,\lambda,\tau,T,\omega_0)$.

\subsection{Eckhaus instability}
Eckhaus instability is characterized by a change in the curvature of $\lambda(\phi)$ at $\phi=0$ \cite{hoyle2006pattern}. We follow in spirit the treatment of the Eckhaus instability developed for spatially continuous, partial differential equations \cite{Rademacher2007,hasan2021spatiotemporal}. Accordingly, we compute the first and second derivatives $\lambda_\phi$ and $\lambda_{\phi\phi}$ at $\phi=0$. This can be achieved by differentiating condition~(\ref{eq:BVP2}) with respect to $\phi$, noting that only $\lambda$ and $\zeta$ depend on $\phi$, and augmenting the system with appropriate normalization and phase conditions.

The first derivative satisfies
\begin{equation}\label{eq:BVPe1}
\begin{aligned}
\zeta_\phi^\prime&(t) = T\!\left(\!-\lambda_\phi \zeta(t) - \lambda \zeta_\phi(t) +  \right. \\
& \left.\!\!\!\!\!\!\!\sum_{-1\leq m \leq 1 } \!\!\!\!\!  e^{im\phi} B_m(t) \left(im\zeta\left(t-m\frac{\tau}{T}\right)+\zeta_\phi\left(t-m\frac{\tau}{T}\right)\right)\!\right),
\end{aligned}
\end{equation}
together with the normalization, phase and periodic conditions
\begin{align}
    0=& \int_0^T \langle \zeta, \zeta_\phi\rangle dt, \label{eq:BVPe2}\\
    0=& \text{Im}(\pi_1(\zeta_\phi(0))),  \label{eq:BVPe3}\\
    \zeta_\phi(0)=&\zeta_\phi(1) , \label{eq:BVPe4}
\end{align}
which, combined with (\ref{eq:BVP1})–(\ref{eq:BVP8}), define a well-posed 2PBVP for $(\zeta_\phi,\lambda_\phi)$.

Similarly, the second derivative satisfies
\begin{equation}\label{eq:BVPee1}
\begin{aligned}
\zeta_{\phi\phi}^\prime(t) = &T\!\left(\!-\lambda_{\phi\phi} \zeta(t) - 2\lambda_\phi \zeta_\phi(t) -\lambda\zeta_{\phi\phi}(t)\right) + \\
&T\sum_{-1\leq m \leq 1 } \!\!\!\!\!  e^{im\phi} B_m(t) \left(-m^2\zeta\left(t-m\frac{\tau}{T}\right) +  \right. \\
& \qquad \left. 2im \zeta_\phi\left(t-m\frac{\tau}{T}\right) + \zeta_{\phi\phi}\left(t-m\frac{\tau}{T}\right)\!\right), 
\end{aligned}
\end{equation}
with corresponding normalization and phase conditions
\begin{align}
    0=& \int_0^T \langle \zeta, \zeta_{\phi \phi}\rangle dt, \label{eq:BVPee2}\\
    0=& \text{Im}(\pi_1(\zeta_{\phi\phi}(0))), \label{eq:BVPee3} \\
    \zeta_{\phi\phi}(0)=&\zeta_{\phi\phi}(1) , \label{eq:BVPee4}
\end{align}
which, together with (\ref{eq:BVP1})–(\ref{eq:BVP8}) and (\ref{eq:BVPe1})–(\ref{eq:BVPe4}), define a well-posed 2PBVP for $(\zeta_{\phi\phi},\lambda_{\phi\phi})$. We remark that, when considering the full 2PBVP formulation for the curvature, condition~\eqref{eq:BVP5} changes to
$$0 = \int_0^1 \langle \tilde{z}, \tilde{z}'_{\text{ref}}\rangle + \langle \zeta, \zeta'_{\text{ref}}\rangle+ \langle \zeta_\phi, \zeta'_{\phi \,\text{ref}}\rangle+ \langle \zeta, \zeta'_{\phi\phi \, \text{ref}}\rangle\,dt. $$

Fixing $\phi=0$, we solve the extended 2PBVP (\ref{eq:BVP1})–(\ref{eq:BVP8}) and (\ref{eq:BVPe1})–(\ref{eq:BVPee4}) while allowing $(\tilde z,\zeta,\zeta_\phi,\zeta_{\phi\phi},\lambda,\lambda_\phi,\lambda_{\phi\phi},k,\tau,T)$ to vary until $\mathrm{Re}\lambda_{\phi\phi}=0$. We then fix $\mathrm{Re}\lambda_{\phi\phi}$ and continue in $\alpha$, using $(\tilde z,\zeta,\zeta_\phi,\zeta_{\phi\phi},\lambda,\lambda_\phi,\text{Im}\lambda_{\phi\phi},k,\alpha,\tau,T)$ as free variables. In this way, the resulting set of points $(\alpha,k)$ defines the Eckhaus boundary shown in Fig.~\ref{fig:eckhaus-curve}. 

\section{Conclusions}
\label{sec:conclusions}

We investigated the existence and stability of nonlinear traveling waves in lattices of mechanical oscillators with odd elasticity (nonreciprocal coupling). Through direct numerical simulations and numerical continuation, and by using the master stability framework \cite{RuschelGiraldo}, we demonstrated the emergence of both periodic and quasiperiodic traveling waves. This motivated the introduction of an analysis based on the rotating-wave approximation (RWA) to describe the slow envelope dynamics of these traveling waves.

We analyzed the homogeneous (static) equilibria and characterized the onset of oscillatory instabilities through linear and bifurcation analysis. Building on this framework, we studied the existence and stability of periodic traveling waves and showed that their destabilization is governed by an Eckhaus-type instability. In particular, the master stability framework enabled a quantitative characterization of size effects, including the determination of stability bounds for different finite lattice embeddings and the appearance of sporadic stable sizes.

We showed that the RWA accurately captures the dynamics of the periodic traveling waves near their onset. In particular, the RWA reproduces key spectral properties of the traveling waves and provides a good approximation of the corresponding Eckhaus instability boundaries. Beyond the regime of stable periodic traveling waves, we identified quasiperiodic solutions and demonstrated strong agreement between direct numerical simulations of the full finite lattice model and numerical continuation based on the RWA. Finally, we computed existence boundaries for quasiperiodic traveling waves, revealing how localization and stability depend on the interplay between nonlinearity, dissipation, and odd elasticity.

An important direction for future work concerns a detailed analysis of the quasiperiodic regime, which appears considerably more challenging than the analysis of periodic traveling waves. In particular, the quasiperiodic states exhibit complex transitions, localization phenomena, and stability changes that are not yet fully understood. The rotating-wave approximation provides a promising reduced framework for investigating these dynamics, opening the possibility of systematically studying how quasiperiodic motion emerges in nonreciprocal mechanical lattices through sequences of local and global bifurcations. A more exhaustive bifurcation analysis of the rotating-wave approximation, including the organization of invariant tori and secondary instabilities, is therefore a natural and important direction for future research. 

Our findings can inform novel designs of active mechanical systems for directional control of energy transport. More broadly, our results establish the rotating-wave approximation and master stability framework as effective tools for analyzing nonlinear wave phenomena in nonreciprocal finite lattices, opening the door to the systematic understanding of more complex coherent structures and higher-dimensional lattices with odd elasticity. 

\section*{Acknowledgments}
A.G. was supported by KIAS Individual Grant No. CG086102 at Korea Institute for Advanced Study. S.R. was supported by UKRI Grant No. EP/Y027531/1. B.Y. acknowledges the Natural Sciences and Engineering Research Council of Canada for supporting a research visit to KIAS. The code and the data that support the findings of this study are openly available at the following URL/DOI:
\url{https://github.com/andrusgiraldo/GRY_NonReciprocalMechanic}.


\bibliographystyle{unsrt} 
\bibliography{GRY.bib}

\end{document}